\documentclass[reprint,aps,prx,twocolumn,superscriptaddress,amsmath,amssymb,floatfix]{revtex4-2}
\usepackage{graphicx}
\usepackage{layout}
\usepackage{natbib}
\usepackage{dcolumn}
\usepackage{braket}
\usepackage{bm}
\usepackage{color}
\usepackage{verbatim}
\usepackage{hyperref}

\newcommand{\ale}[1]{{#1}}

\begin{document}

\title{Anomalous periodicity and parafermion hybridization in superconducting qubits}

\author{Alessio Calzona}
\affiliation{Institute for Theoretical Physics and Astrophysics, University of W\"urzburg, 97074 W\"urzburg, Germany}

\author{Matteo Carrega}
\affiliation{CNR-SPIN, Via Dodecaneso 33, 16146 Genova, Italy}

\author{Luca Chirolli}
\affiliation{Department of Physics, University of California Berkeley, CA-94720, USA}
\affiliation{Istituto Nanoscienze - CNR, I-56127 Pisa, Italy}

\begin{abstract}
Topological quantum computation relies on a protected degenerate subspace enabling complicated operations in a noise-resilient way. To this end, hybrid platforms based on non-Abelian quasiparticles such as parafermions hold great promise. These are predicted to emerge at the interface between fractional quantum Hall states and superconductors and therefore naturally couple to superconducting qubits. Here, we study a parafermionic fluxonium circuit and show that the presence of topological states yields a striking periodicity in the qubit spectrum. In addition, peculiar and marked signatures of different parafermion coupling, associated with multiple tunneling of fractional quasiparticles, can be detected in the qubit microwave spectrum. Finite parafermion coupling can reduce the full degeneracy of the non-Abelian manifold, and we show that this configuration can be used to assess the remaining degree of protection of the system. 
\end{abstract}

\maketitle

{\it Introduction.---}
The possible existence of non-Abelian quasiparticles~\cite{nayak2008, stern2008}, together with their potential in performing non-trivial unitary operations on a degenerate subspace, have triggered a great scientific effort recently. The ensuing topological protection against local perturbations holds great promise for noise-resilient quantum computation~\cite{ladd2010, devoret2013, arute2019, haldane2017, stern2013} and non-Abelian statistics constitutes a first key element toward fault-tolerant quantum computation~\cite{nayak2008, stern2013}. Novel topological solid-state platforms range from exotic fractional quantum Hall (FQH) states  \cite{nayak2008, stern2008, carrega2021, nakamura2020, bartolomei2020, willett2019} to hybrid superconductor-semiconductor devices~\cite{prada2020, mourik2012, marcus2020, matsuo2021, kurtossi2021} and topological insulators~\cite{bocquillon2018, hajer2019}. These may host Majorana zero-energy modes (MZMs)~\cite{prada2020, oreg2010, kitaev2001} or new topological states of matter called parafermions (PFs) (or fractionalized Majoranas) \cite{fendley2012, schmidt2015, alicea2016, barkeshli2014, mong2016, liang2019, cobanera2013, burnell2016, calzona2018, rossini2019, ziani2019, mazza2018, schmidt2020r, ronetti2021}. However, despite the huge ongoing experimental efforts, up-to-date very scarce evidence of non-Abelian statistics in FQH systems has been reported \cite{carrega2021, willett2019}, together with no unambiguous detection of MZMs~\cite{prada2020, pan2020, zhang2021}. 

Parafermions can be thought of as a non-trivial generalization of MZMs featuring, for instance, a larger topological degeneracy, which results in greater computational capabilities 
\cite{hutter2016}. Moreover, differently from MZMs, PFs can be coupled via several different and intriguing mechanisms \cite{teixeira2022}. A full characterization of these couplings, which is still lacking to the best of our knowledge, would represent a precious tool to detect the presence of PFs, manipulate them and clarify their interplay with the system they are embedded in \cite{burnell2016,schmidt2019}. In this respect, one of the most promising platforms for hosting parafermions is a FQH state in proximity to a superconductor (SC) characterized by strong spin-orbit interactions. Indeed, the strong correlations present in the system \cite{wan2015, guiducci2019, amet2016, lee2017, zhao2020, gul2022} allow for the localization of these entangled zero-energy modes at the SC-FQH interface \cite{lindner2012,clarke2013,clarke2014,katzir2020}. Different possible realizations have been theoretically inspected, with various FQH filling factors or SCs geometry~\cite{lindner2012, clarke2013, clarke2014, schmidt2015, katzir2020, schmidt2019, nielsen2022}, but still await any experimental evidence.   

Here, we study spectroscopic signatures of PFs and their hybridization, within a setup consisting of a FQH-based platform and a fluxonium superconducting qubit~\cite{manucharyan2009}. In particular, we show how superconducting circuits~\cite{devoret2013, bell2016, mencia2020, somorof2021, larsen2020}, which have been already proposed as efficient tools to detect MZMs~\cite{hassler2010,jiang2011,bonderson2011,hou2011,hassler2011,pekker2013,ginossar2014,pikulin2019,yavilberg2019,avila2020superconducting,avila2020majorana,chirolli2022}, can be also implemented for the detection of PFs and, crucially, for the direct characterization of non-trivial PF hybridization, absent in the MZM case, efficiently discriminating between different coupling mechanisms. More in detail, we consider $\mathbb{Z}_{2m}$ PFs, hosted by a hybrid SC-FQH system in the $\nu=1/m$ Laughlin state. In analogy with a specific proposal for the detection of MZMs~\cite{pekker2013}, we consider a fluxonium qubit sharing the SC leads with the hybrid setup and study its coupling with the PFs. We show how the PF-mediated $4m\pi$-Josephson effect yields a $4m\pi$ periodicity of the qubit spectrum as a function of the applied external flux. In addition, we consider several terms describing the hybridization of the four PFs in the circuit. These extra spurious terms, which have never been \ale{explicitly} considered in the literature to our knowledge, modify the periodicity of the spectrum in a very recognizable way, thus providing a striking signature of the presence of PFs and of their properties. We discuss the physical origin of these couplings and propose possible ways to enhance or suppress their action through local gating on the QH region, i.e. using a tip of a scanning gate microscope (SGM)\cite{leroy2005, jalabert2010, jura2009, paradiso2010, paradiso2012, bours2017, nowak2022}. As a relevant observable, we study the microwave (MW) spectrum detected by a nearby MW resonator and show how clear evidence of the $4m\pi$ periodicity appears, together with marked specific signatures of PF coupling through anomalous periodicity. Our results open the way to a systematic study of parafermionic topological state and of fundamental quasiparticle processes in the system, that complement well proposals based on transport spectroscopy \cite{nielsen2022}.

\begin{figure}[t]
\includegraphics[width=\linewidth]{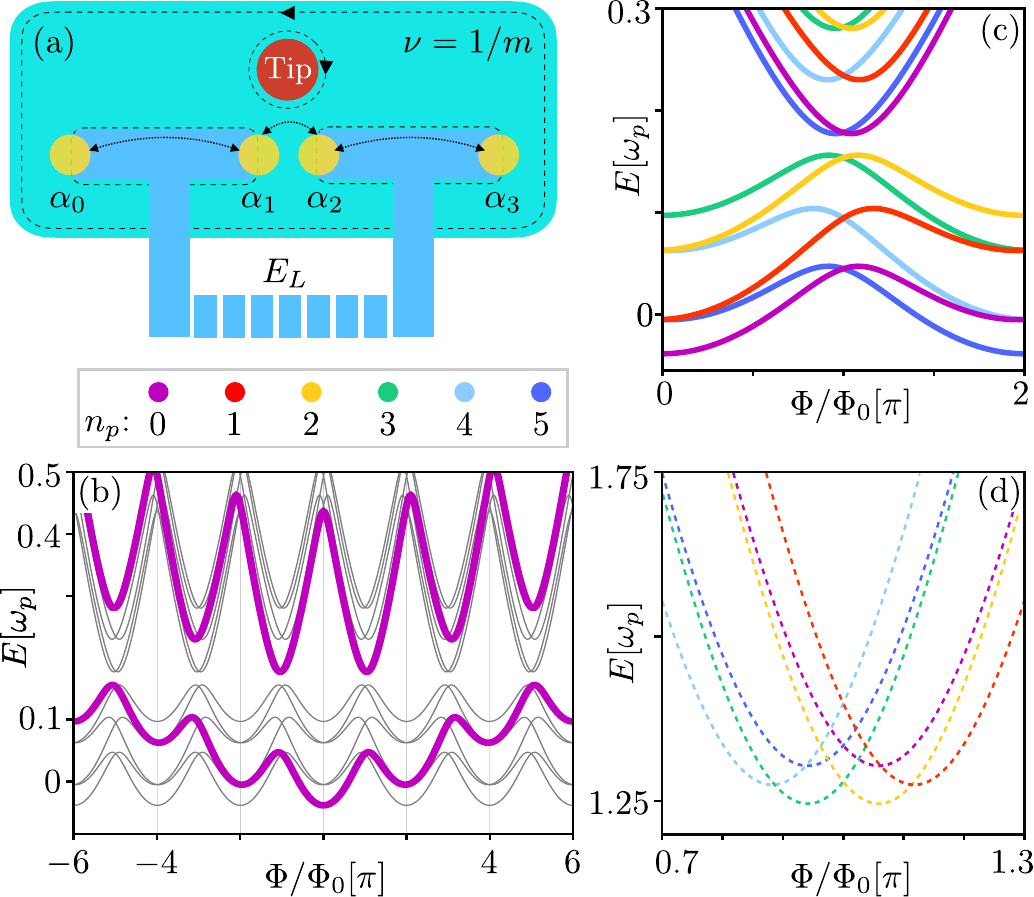}
\caption{(a) Schematics of a parafermionic fluxonium. Four PFs $\alpha_0,\ldots,\alpha_3$ (yellow) localize at the ends of two SC leads (blue), shunted by a superinductor. (b) $12\pi$-periodic spectrum for $m=3$, showing the ground and first excited fluxonium states for the 6 possible PF numbers $n_p$ (thick purple line corresponds to $n_p=0$). (c) Zoom of (b) around $\Phi/\Phi_0=\pi$. (d) MW spectrum showing allowed transitions assuming all $n_p$ states are equally populated. Transitions belonging to a given $n_p$ are marked with a color-coded dashed line. Parameters are $E_J=0.45\,\omega_p,\,E_L=0.03\,\omega_p,\gamma^{(1)}=0.04\,\omega_p,\gamma^{(2,3)}=\lambda^{(k)}=0$.
}
\label{Fig1}
\end{figure}

{\it Parafermion system.---}
We consider the system shown in Fig.~\ref{Fig1}(a), describing a FQH system in the $\nu=1/m$ Laughlin state, interrupted at the center by two trenches that host superconducting leads, such as NbN \cite{lee2017}. Four parafermions $\alpha_i$ are predicted to emerge at the ends of the two SC leads, with $i=0,\ldots,3$. The localization of PFs is well described in Refs.~\cite{clarke2013, schmidt2020r}: chiral FQH edge states of given spin projection are brought into proximity on the sides of the SC leads, as schematized in Fig.~\ref{Fig1}(a), and effectively form a Luttinger liquid \cite{Note1}. The latter is specified by two independent non-chiral fields, $\phi$ and $\theta$,  obeying $[\phi(x),\theta(x')]=i(\pi/m)\Theta(x-x')$ with $\rho=\partial_x\theta/\pi$ the electronic density. The relevant pairing acting underneath the SC leads generates an inhomogeneous sine-Gordon term that pins $\phi$ to one of the $2m$ inequivalent minima of the potential, $\phi=\pi \hat{N}_\mu/m$, with $N_\mu=0,\ldots, 2m-1$. Analogously, outside the SC regions, the Hall droplet induces strong backscattering between the chiral modes that fixes $\theta$ and we can then set $\theta=\pi\hat{N}_c/m$ and $\theta=\pi \hat{N}^{l(r)}_b/m$ in the central, left and right regions, respectively. Commutation relations between $\phi$ and $\theta$ fields induce $[\hat{N}_R,\hat{N}_c]=im/\pi$ and $[\hat{N}_{L(R)},\hat{N}^l_b]=im/\pi$, and localized modes appear at the ends of the SC leads, that are written as \cite{clarke2013}
\begin{equation}
\alpha_{1(2)}=e^{i\frac{\pi}{m}(\hat{N}_{L(R)}+\hat{N}_c)},~ \alpha_{0(3)}=e^{i\frac{\pi}{m}(\hat{N}_{L(R)}+\hat{N}^{l(r)}_b)}.
\end{equation}

The resulting $\mathbb{Z}_p$ parafermions with $p=2m$ are characterized by peculiar algebraic properties, $\alpha_i ^ p = 1$, $\alpha_i ^ {p-1}  =\alpha_i^\dagger$, and $\alpha_i \alpha_j = e^{i2\pi/p }\, \alpha_j \alpha_i$ (for $i<j$), generalizing the Majorana case $p=2$ ($m=1$). The Fock space associated with a single pair of $\mathbb{Z}_p$ PFs is $p$-dimensional. \ale{In particular, since $e^{i \pi/p} \alpha_1^\dagger \alpha_2 = e^{2 \pi i (\hat{N}_R-\hat{N}_L)/p}$, the state of the pair $\{\alpha_1,\alpha_2\}$ can be specified by the number} $\hat{N}_R-\hat{N}_L$ \cite{Note2}. \ale{The conservation of the total PF parity $\hat{P}=\alpha^\dag_1\alpha_2\alpha^\dag_3\alpha_0= e^{2\pi i/p(\hat{N}_b^l-\hat{N}_b^r)}$ allows us to identify $p$ decoupled sectors, with fixed total PF parity, within the whole $p^2$-dimensional space associated with the four PFs $\alpha_0,\dots,\alpha_{3}$.} It is useful to introduce the so-called ``Fock parafermions" \cite{cobanera2014, rossini2019, schmidt2019}, which allow defining number operators $\hat{n}$ and to properly label the states spanning pairs of PFs. Operatively, we can set $\hat{n}_p=(\hat{N}_L-\hat{N}_R-1)$ mod$(p)$ and  $\hat{n}_0+\hat{n}_p=(\hat{N}_b^l-\hat{N}_b^r-1)$ mod$(p)$  (for details, see \cite{Note1}).

The coupling between the two PFs $\{\alpha_1,\alpha_2\}$ localized at the junction accounts for  tunneling of quasiparticles with fractional charge $2e/p$ between the two SC leads. This can be seen by noticing that the operator $\hat{U}=\alpha_2^\dag\alpha_1$ yields $\hat{U}^\dag \hat{N}_c\hat{U}=\hat{N}_c+1$ \cite{clarke2013}. Single Cooper pair tunneling between the SC leads requires then $p$ iterations to complete a cycle, rendering the energy $2p\pi $-periodic in the gauge invariant phase difference $\varphi$ between the SC leads. A closer inspection reveals that this is not the only possible process at the junction and that multiples of charge $e/m$ can tunnel through the gapped region between the SC leads.  Indeed, powers of $\hat{U}$ increase the charge in the central region in the form $(\hat{U}^\dag)^k\hat{N}_c\hat{U}^k=\hat{N}_c+k$. This corresponds to $2\pi k$ slippage of the field $\theta$, from one minimum to the $k$-th nearest minima of the cosine potential \cite{Note1}. By artificially reducing backscattering between the fractional edge channels in the central region between the SC leads, we can allow for higher order processes and asymptotically approach the $2p\pi$-periodic sawtooth current-phase relation of the fractional Josephson effect in highly transparent junctions \cite{clarke2013}. This can be done via local gating in the FQH region through the tip of a local SGM \cite{jura2009, paradiso2010, paradiso2012, bours2017} that produces an antidot and enhances resonant tunneling between the fractional Hall edge states, as depicted in Fig.~\ref{Fig1}(a). At lowest orders, the most general PF coupling at the junction reads
\begin{equation}\label{Eq:H0PF}
	H^{(0)}_{\rm PF} = \sum_{k=1}^{p/2}\gamma^{(k)} \cos\left(k\frac{\varphi+2\pi \hat{n}_p}{p} \right),
\end{equation}
that is controlled by the amplitudes $\gamma^{(k)}$. The latter depend exponentially on the distance between the PFs compared to the magnetic length and can be calculated through the instanton technique \cite{coleman1985,burnell2016,teixeira2022,schmidt2019}. In the perturbative regime, tunneling of $ke/m$ quasiparticles is suppressed as $\gamma^{(k)}\propto \gamma^k$.

\begin{figure}[t]
\includegraphics[width=\linewidth]{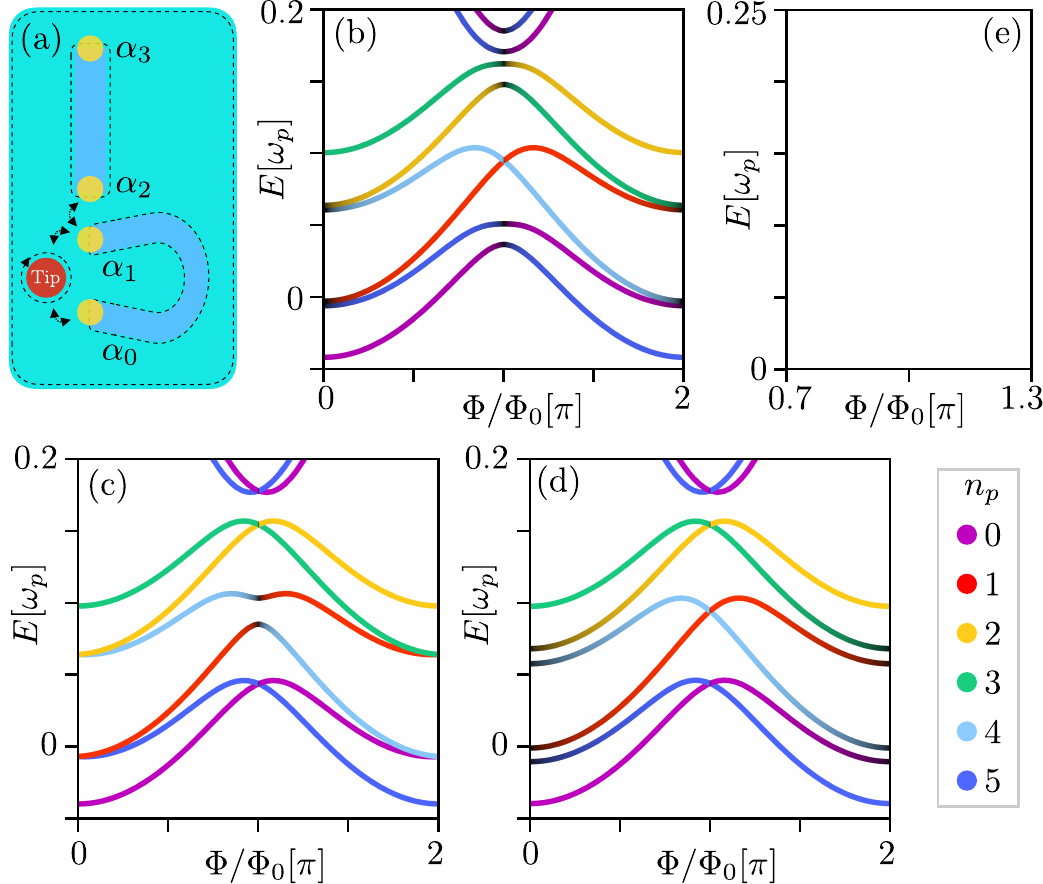}
\caption{(a) Schematics of the setup with a U-shaped SC lead and a local tip that allows for tuning of the couplings $\lambda^{(k)}$. (b,c,d) Spectrum of the lowest fluxonium band in presence of off-diagonal PF couplings. Black coloring of the lines indicates hybridization between different $n_p$ values. In (b), $\lambda^{(1)}=0.015\omega_p$ lifts all crossings, giving $2\pi$-periodicity. In (c), $\lambda^{(2)}= 0.01\omega_p$ describes charge $2e/3$ tunneling, giving $4\pi$-periodicity. In (d), $\lambda^{(3)} = 0.01\omega_p$ describes charge $e$ tunneling, giving $6\pi$-periodicity. All the other parameters are identical to Fig.~\ref{Fig1}. (e) MW spectrum corresponding to the scenario of panel (b), when only the ground state is populated.}
\label{Fig2}
\end{figure}

We can proceed analogously and consider tunneling of quasiparticles through the SC leads. \ale{Focusing on the left SC lead and considering only the $p$-dimensional sector spanned by $\hat{n}_p$ for a fixed total PF parity, the generic coupling of PFs $\{\alpha_0,\alpha_1\}$ reads} \cite{Note1}
\begin{equation}\label{Eq:H1PF}
	H^{(1)}_{\rm PF} = \sum_{k=1}^{p/2} \lambda^{(k)} \alpha_1^k + {\rm H.c.}.
\end{equation}
The amplitudes $\lambda^{(k)}$, associated with $2k\pi$ phase slippage of the field $\phi$ and charge $ke/m$ quasiparticle tunneling, are controlled by the coherence length of the superconductor and, in the perturbative regime, they scale as $\lambda^{(k)}\propto\lambda^k$. \ale{To actively manipulate those amplitudes, we envision a slightly modified geometry, depicted in Fig.~\ref{Fig2}(a), where a U-shaped left SC lead allows for direct PF coupling via the quantum Hall gap, controlled by local gating~\cite{shabani_nanolett,paradiso2010, paradiso2012}. This opens the possibility to study several fundamental processes of quasiparticle tunneling and asses their effects on the parafermionic spectra. Their detection and characterization, enabled by the fluxonium qubit described below, are of great relevance as they non-trivially extend the physics of Majorana hybridization, which is inherently limited to electronic tunneling.}

{\it Parafermionic fluxonium response.---}
\ale{The coupling of the parafermion system to a superconducting fluxonium qubit \cite{somorof2021} realizes an interesting conventional/topological architecture~\cite{chirolli2022} and is obtained by closing the Josephson junction between the left and right SC leads on a superinductor, formed by a long chain of Josephson junctions.}  
The fluxonium Hamiltonian reads  
\begin{equation}
	H_F = -4 E_C \partial_\varphi^2 + \frac{1}{2} E_L (\varphi-2\pi\Phi/\Phi_0)^2-E_J \cos(\varphi),
\end{equation}
where $E_J$ is the Josephson energy~\cite{noteEJ}, $E_L = (\Phi_0/2\pi)^2/L$ is the inductive energy, $E_C = e^2/(2C)$ is the charging energy and the external flux $\Phi$ is measured in units of $\Phi_0/(2\pi) = \hbar/(2e)$. PFs $\alpha_i$ couple to the fluxonium through the $\varphi$-dependent term Eq.~(\ref{Eq:H0PF}) and the total Hamiltonian including PF couplings Eq.~(\ref{Eq:H1PF}) reads $H = H_F+H^{(0)}_{\rm PF}+H^{(1)}_{\rm PF}$.

An effective Hamiltonian can be derived by introducing fluxon states $|n_\varphi\rangle$, which describe states of circulating current in the loop whose number changes via tunneling of fluxoids at the Josephson junction described by the term $T^\pm_\varphi|n_\varphi\rangle=|n_\varphi\pm1\rangle$. In addition, we consider eigenstates $|n_p\rangle$ of $\hat{n}_p$, through which the PF coupling at the junction reads $\gamma^{(k)}\cos(2k\pi(n_\varphi+n_p)/p)$.
Finite PF coupling in the SC leads introduce coupling between different $|n_p\rangle$ states through generalized Pauli matrices.  In the combined basis $|n_\varphi,n_p\rangle$ the low-energy Hamiltonian is
\begin{eqnarray}\label{Eq:HFlowEn}
H_{\rm eff}&=&\frac{E_L}{2}(2\pi n_\varphi-2\pi\Phi/\Phi_0)^2-\frac{E_S}{2}\sum_{a=\pm}T^a_\varphi\nonumber\\
&+&\sum_{k=2}^{p/2}\gamma^{(k)}\cos(2\pi k(n_\varphi+n_p)/p)+H_{\rm PF}^{(1)}.
\end{eqnarray}
with $E_S=E_S(E_C,E_J,\gamma^{(1)})$ the phase slip rate at the Josephson junction \cite{mooij2006,koch2009}. 
The spectrum of the Hamiltonian is obtained by numerical diagonalization of $H$ as a function of the external flux $\Phi$. For definiteness, we choose the simplest $m=3$ case ($\mathbb{Z}_6$ PFs), set the parameter regime $\ale{0<\gamma^{(1)}}<\pi^2E_L<E_J$ \cite{pekker2013}, and introduce the plasma frequency $\omega_p=\sqrt{8 E_C E_J}$. The spectrum features two bands, which originate from groups of parabolic curves spaced in $\Phi/\Phi_0$ approximately by $2\pi$ with anticrossings around $\Phi/\Phi_0\sim\pi+2n\pi$ of size $E_S$, in agreement with Eq.~(\ref{Eq:HFlowEn}).

\ale{At first, we consider all $\lambda^{(k)} = 0$, so that PF tunneling only occurs at the junction and $\hat{n}_p$ is a conserved quantity. The result is shown in Fig.~\ref{Fig1}(b), where we only consider $\gamma^{(1)}\neq 0$. Within each band, we observe $p=6$ different curves corresponding to the possible PF occupation numbers $n_p=0,\dots 5$.} Each curve exhibits a $12\pi$ periodicity. In Fig.~\ref{Fig1}(c), we focus on the interval $[0,2\pi]$ to highlight the presence of protected crossings between curves associated with different values of $n_p$.  We observe crossings at $\Phi/\Phi_0=2n\pi$, involving states whose PF number differs by $2$ mod$(p)$, and at $\Phi=\pi+2n\pi$, involving states which differ by $1$ or $3$ mod$(p)$. Additional PF couplings $\gamma^{(2)}$ and $\gamma^{(3)}$ at the junction do not change the $12\pi$ periodicity and, therefore, do not qualitatively modify our findings~\cite{Note1}. This is an important result of the present work: the spectrum of the fluxonium allows for detection of the $2p\pi$-Josephson effect, which is not affected by the presence of extra tunneling processes through the junction.

A useful tool to characterize the spectral properties of the system is provided by the microwave spectrum (MWS) that is read out by a nearby resonator. Assuming a minimal inductive coupling to the phase operator $\hat{\varphi}$, the MWS captures transitions from the ground state $|0\rangle$ to the excited states $|n\rangle$ and it is given by Fermi golden rule expression $S(\omega)=\sum_n\left|\langle 0|\hat{\varphi}|n\rangle\right|^2\delta(\omega_n-\omega_0-\omega)$. In Fig.~\ref{Fig1}(d), we show the MWS around $\Phi/\Phi_0 \sim \pi$ obtained by considering transitions from each one of the $6$ lowest energy states, with fixed $n_p=0,\dots,5$, to the corresponding first excited states. Depending on $n_p$ (see the color of the dashed lines), the position of the minimum is shifted from $\Phi/\Phi_0=\pi$. The existence of these different minima and the presence of protected crossings in Fig.~\ref{Fig1}(d) are a good (albeit indirect) indication of the system $12\pi$ periodicity. The latter can be directly observed by preparing the system in a single eigenstate of $\hat{n}_p$. In this case, a sweep of $\Phi/\Phi_0$ from $-6\pi$ to $6\pi$ reveals indeed all the $6$ inequivalent minima located around $\Phi/\Phi_0\sim\pi+2m\pi$ (see \cite{Note1}). In passing, we note that a (possibly time-resolved) microwave spectroscopy can be used to investigate population transfer between sectors with different $n_p$, due to external mechanisms, by studying changes in the relative intensity of the different minima in the MWS. This can shine a light on the role played by different kinds of quasiparticle poisoning events whose understanding, in analogy with the Majorana case \cite{goldstein2011,rainis2012,budich2012,karzig2021}, is of great relevance, e.g., for the future development of PF-based qubits.

\begin{figure}[t]
\includegraphics[width=\linewidth]{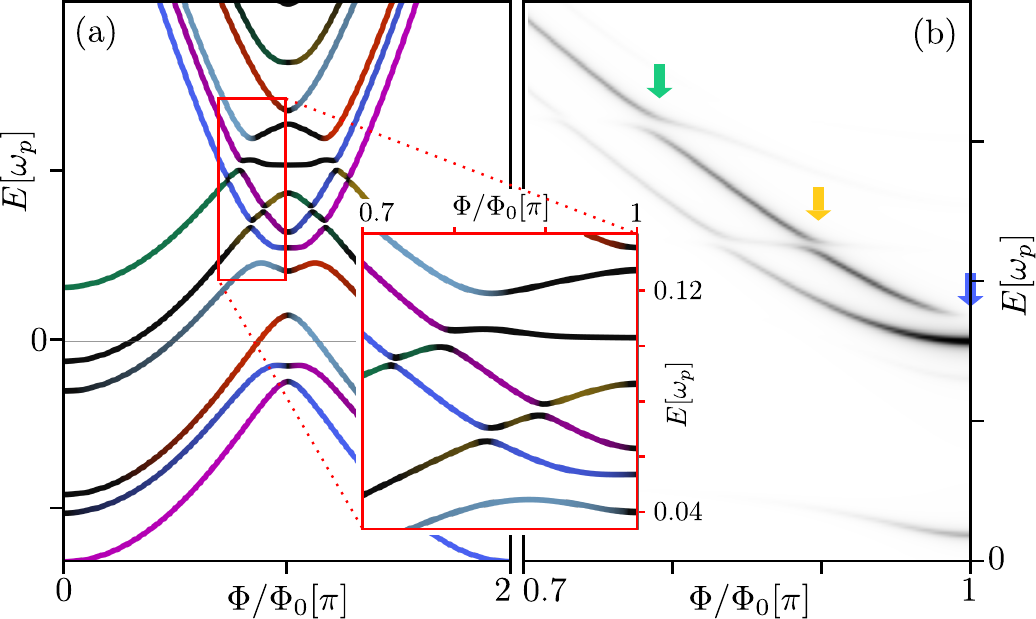}
\caption{(a) Spectrum of the device for $E_L=0.03\,\omega_p,\,E_J=0.5\,\omega_p,\,\gamma=0.05\,\omega_p$, and $\lambda^{(1)}=\lambda^{(2)}=\lambda^{(3)}=0.02\omega_p$. Inset: zoom showing the avoided crossings. (b) MWS obtained by populating only the ground state. \label{Fig3}}
\end{figure}

We now study the effect of PF coupling within the left SC lead, i.e. of finite $\lambda^{(k)}$. Since a non-zero $H^{(1)}_{\rm PF}$ does not conserve $n_p$, the degeneracy at the crossing points is generally lifted by the off-diagonal PF couplings. Importantly, however, depending on which type of coupling is considered, the crossings are lifted in different ways. If the two PFs on the left SC lead exchange charge $ke/3$ quasiparticles, the splitting of the crossings occurs for states whose $n_p$ differ by $\bar mk$ mod$(p)$, where the integer $\bar m\geq 1$ indicates the order of the process in $\lambda^{(k)}$. In particular, for $\lambda^{(1)} \neq 0$, all the crossings are split, with the largest gaps appearing between states whose $n_p$ differ by $1$. This scenario is shown in Fig.~\ref{Fig2}(b), where the overall periodicity is reduced from $12\pi$ to $2\pi$ and two large splittings appear at $\Phi/\Phi_0=\pi$ between states with $n_p=5,1$ and $n_p=2,3$. Second order processes open smaller energy gaps at $\Phi/\Phi_0=2m\pi$ while a third order process determines an even smaller anticrossing between the $n_p=1$ and $4$ states. Analogously, the (exclusive) presence of a non-zero $\lambda^{(2)}$ or $\lambda^{(3)}$ does not open all the crossings. Tunneling of charge $2e/3$ quasiparticles for finite $\lambda^{(2)}$ only conserves $n_p~{\rm mod}(2)$ and the periodicity of the spectrum is reduced to $4\pi$, as shown in Fig.~\ref{Fig2}(c). Finally, tunneling of charge $e$ quasiparticles for finite $\lambda^{(3)}$ conserves $n_p{\rm mod}(3)$ and the periodicity of the spectrum is reduced to $6\pi$, see Fig. 2(d). In general, we have shown that $ke/3$ tunneling conserves the PF number $n_p$ mod$(k)$, leading to $2k\pi$-periodicity of the spectrum of the parafermionic fluxonium. The analysis of the latter can thus precisely detect the presence of different tunneling mechanisms, also characterizing their strength and possible tunability via local gating. This represents a second important result of our work, which paves the way for the identification of different types of PF hybridizations in a hybrid FQH setup, distinguishing them in terms of their effects on the protection of the topological manifold of the PF system.

The effects of non-zero $\lambda^{(k)}$ also emerge in the MWS. In Fig.~\ref{Fig2}(e), for example, we show the MWS corresponding to the spectrum shown in panel (b), characterized by $\lambda^{(1)}\neq 0$, for the case in which only the ground state is populated. Three different transitions are activated, which directly signals the presence of tunneling of charge $e/3$ quasiparticle, and thus the hybridization of states whose $n_p$ differ by $1$. In particular, a faint low energy transition is activated between the ground state and the first excited state, resulting from the hybridization of the $n_p=0,5$ PF states. The more pronounced second and third transitions involve the excited states of the fluxonium degree of freedom, still featuring $n_p=0,5$. To appreciate even better the effects of $\lambda^{(k)}$ on the MWS, we slightly modify the parameters of the fluxonium by increasing $E_J$, so that some states belonging to the second fluxonium band intersect with the ones of the first band. These additional crossings make it straightforward to detect and distinguish the presence of different non-zero $\lambda^{(k)}$ by analyzing the MWS. We demonstrate this in Fig.~\ref{Fig3} where, to show the maximal effect of PF hybridization, we simultaneously switch all $\lambda^{(k)}$ on. The fluxonium spectrum, in panel (a), features several avoided crossings that emerge also in the corresponding MWS, shown in panel (b), obtained by only populating the ground state. There, the main gaps opened by $\lambda^{(1)},\,\lambda^{(2)}\text{ and }\lambda^{(3)}$ (highlighted with blue, yellow, and green arrows, respectively) can be easily identified and distinguished between each other.

{\it Conclusions.---} We have studied a parafermionic fluxonium circuit, where signatures of PF hybridization can be clearly detected with MWS, identifying the degree of topological degeneracy. The elucidated ``selection" rules hold also in presence of additional crossings, which can result from the touching of the two fluxonium bands around $\Phi/\Phi_0\sim\pm\,\pi,\,\dots$ in presence of a larger $E_J$. In this respect, we note that one could also envision more complicated couplings that involve three or four PFs, while still conserving the total PF parity. Those terms would result in the generalization of the $\sigma_y$ operator that may appear in the Majorana case. However, we expect the ``selection" rules highlighted so far to still be valid. 

\begin{acknowledgments}
A.C. was supported by the W\"{u}rzburg-Dresden Cluster of Excellence ct.qmat, EXC2147, project-id 390858490, and the DFG (SFB 1170). A.C. also thanks the Bavarian Ministry of Economic Affairs, Regional Development and Energy for financial support within the High-Tech Agenda Project ``Bausteine f\"{u}r das Quanten Computing auf Basis topologischer Materialen''. L. C. has received funding from the European Union’s Horizon 2020 research and innovation programme under the Marie Sklodowska-Curie grant agreement No 841894.
\end{acknowledgments}


\pagebreak
\begin{widetext}
\begin{center}
	\textbf{\large Supplemental material for ``Anomalous periodicity and Parafermions hybridization in superconducting qubits''}
\end{center}
\end{widetext}
\setcounter{equation}{0}
\setcounter{figure}{0}
\setcounter{table}{0}
\setcounter{page}{1}
\makeatletter
\renewcommand{\theequation}{S\arabic{equation}}
\renewcommand{\thefigure}{S\arabic{figure}}
\renewcommand{\bibnumfmt}[1]{[S#1]}
\renewcommand{\citenumfont}[1]{S#1}

	This document contains supplementary information and technical details in support of the results presented in the main text.

\section{Parafermion description}

The low-energy physics and the localization of PFs is well captured by the description provided in Ref.~\cite{Clarke2013, schmidt2020r}, where the major players are chiral counterpropagating FQH edge states of the Laughlin fraction $\nu=1/m$ of given spin projection,  described by the fields $\phi_\mu$, with $\mu={\rm R,L}$, satisfying 
$[\phi_\mu (x),\phi_\mu(x')]=\sigma_\mu i(\pi/m){\rm sgn}(x-x')$, with $\sigma_{\rm R/L}=\pm 1$ and $[\phi_{\rm L}(x),\phi_{\rm R}(x')]=i(\pi/m)$. 
Low-energy right- and left-moving quasiparticles of charge $e/m$ are created by the operators $e^{i\phi_{\rm R/L}}$, that exhibit anyonic exchange statistics, 
\begin{equation}
	e^{i\phi_\mu(x)}e^{i\phi_\mu(x')}=e^{i\phi_\mu(x')}e^{i\phi_\mu(x)}e^{i\sigma_\mu (\pi/m){\rm sgn}(x'-x)},
\end{equation}
from which it follows that the electron operators given by $\psi_{\rm R/L}\propto e^{im\phi_{\rm R/L}}$ obey fermionic statistics. Introducing $\phi_{\rm R/L}=\phi\pm\theta$ obeying $[\phi(x),\theta(x')]=i(\pi/m)\Theta(x-x')$ with $\rho=\partial_x\theta/\pi$ the electronic density, the unperturbed Hamiltonian describing the edge states reads
\begin{equation}
	H_0=\frac{m\nu}{2\pi}\int dx[(\partial_x\phi)^2+(\partial_x\theta)^2].
\end{equation}
Underneath the SC leads the fermionic pairing interaction has the form $\int dx \Delta(x)\psi_L^\dag(x)\psi_R^\dag(x)+{\rm H.c.}$, so that the relevant pairing interaction involves only the $\phi$ field and it is written as
\begin{equation}\label{Eq:PairingInt}
	H_{\rm sc}\propto -\int dx ~\Delta(x)\cos(2m\phi(x)),
\end{equation}
The pairing fixes the field $\phi$ in the superconducting region to one of the $2m$ minima of the cosine, $\phi=\pi \hat{N}_\mu/m$, where the operator $\hat{N}_\mu$ is specified by the eigenvalues $N_\mu=0,\ldots, 2m-1$. Outside the SC regions strong backscattering between the chiral modes is provided by the Hall droplet itself, through an interaction term of the form $\int dx \psi^\dag_R(x)\psi_L(x)+{\rm H.c.}$, that involves only the $\theta$ fields and reads
\begin{equation}
	H_{\rm fqh}\propto -\int dx ~t(x)\cos(2m\theta),
\end{equation} 
that is analogous to that in Eq.~(\ref{Eq:PairingInt}), with $\phi$ replaced by $\theta$ and with $t(x)$ a tunneling amplitude between counterpropagating edge states. The interaction gaps the modes by fixing $\theta$ and we can then set $\theta=\pi\hat{N}_c/m$ and $\theta=\pi \hat{N}^{l(r)}_b/m$ in the central, left and right regions, respectively. The commutation relations satisfied by the fields $\phi$ and $\theta$ induce $[\hat{N}_R,\hat{N}_c]=im/\pi$ and $[\hat{N}_{L(R)},\hat{N}^l_b]=im/\pi$, and localized modes appear at the boundaries of the SC leads, that can be written as \cite{Pekker2013}
\begin{equation}
	\alpha_{1(2)}=e^{i\frac{\pi}{m}(\hat{N}_{L(R)}+\hat{N}_c)},~ \alpha_{0(3)}=e^{i\frac{\pi}{m}(\hat{N}_{L(R)}+\hat{N}^{l(r)}_b)}.
\end{equation}
The resulting $\mathbb{Z}_p$ parafermions with $p=2m$ are characterized by peculiar algebraic properties, $\alpha_i ^ p = 1$, $\alpha_i ^ {p-1}  =\alpha_i^\dagger$, and the anyonic exchange statistics 
\begin{equation}
	\alpha_i \alpha_j = e^{i2\pi/p} \, \alpha_j \alpha_i\quad\text{for } i<j. 
\end{equation}
We now introduce the so-called "Fock parafermions" $d_i$, that allow  to define a number operator and thus to properly label the state. According to Ref.\ \cite{Cobanera2014}, relations connecting an even number of parafermion operators to Fock parafermions that generalize the Majorana case ($p=2$ or equivalently $m=1$) can be written as
\begin{equation}
	\label{eq:PFFock}
	\begin{split}
		\alpha_{2i-1} &= d_i+(d_i^\dagger)^{p-1}\\
		\alpha_{2i} &= e^{i\pi/p}(d_i+(d_i^\dagger)^{p-1})(e^{i2\pi/p})^{N_i} 
	\end{split}
\end{equation}
where the Fock PF number $N_i$ reads
\begin{equation}
	n_i = \sum_{m=1}^{p-1} (d^\dagger_i)^m d_i^m
\end{equation}
that satisfies $[n_i,d_i^\dagger]=d_i^\dagger$ and $[n_i,d_i]=-d_i$. 
The peculiar algebraic properties of parafermions, i.e.
\begin{equation}
	\begin{split}
		&\alpha_i ^ p = 1\\
		&\alpha_i ^ {p-1}  =\alpha_i^\dagger\\	
		&\alpha_i \alpha_j = e^{i2\pi/p} \, \alpha_j \alpha_i\quad\text{for } i<j, 
	\end{split}
\end{equation}
naturally induce the following properties of the Fock PF operators
\begin{equation}
	\begin{split}
		&d_i^p = (d_i^\dagger)^p = 0,\\
		&d_i d_j = e^{i2\pi/p}\, d_jd_i, \quad \text{for }i<j,\\
		&d_i^\dagger d_j = e^{-i2\pi/p}\, d_jd_i^\dagger, \quad \text{for }i<j\\
		&(d_i^\dagger)^m d_i^m + d_i^{p-m}(d_i^\dagger)^{p-m} = 1, \quad \text{for }m=1, \dots p-1.
	\end{split}
\end{equation}
It is important to note that each PF operator $\alpha_{2j-1}$ (and $\alpha_{2j}$) cycles between the $p$ possible occupation numbers associated with the site $j$. Eq.~\eqref{eq:PFFock} can be inverted giving the expression
\begin{equation}
	d_i = \frac{p-1}{p} \alpha_{2i-1} - \frac{1}{p} \sum_{m=1}^{p-1} (e^{i2\pi/p})^{m^2/2 +m}  \alpha_{2i-1}^{m+1}( \alpha_{2i}^\dagger)^{m}.
\end{equation}

In the main text we consider four PF $\alpha_0,\alpha_1,\alpha_2,\alpha_3$, with $\alpha_1$ and $\alpha_2$ located at the two opposite sides of the weak-link of the Josephson junction. The pair $\alpha_1,\alpha_2$ is specified by $\hat{N}_R-\hat{N}_L$, in a way that $e^{i \pi/p} \alpha_1^\dagger \alpha_2 = e^{2 \pi i (\hat{N}_R-\hat{N}_L)/p}$. Analogously we can write $e^{-i \pi/p} \alpha_3^\dagger \alpha_0 = e^{-2 \pi i(\hat{N}_R-\hat{N}_L)/p}e^{2\pi i/p(\hat{N}_b^l-\hat{N}_b^r)}$. The total PF parity can be defined as 
\begin{equation}
	P=\alpha^\dag_1\alpha_2\alpha^\dag_3\alpha_0= e^{2\pi i(\hat{N}_b^l-\hat{N}_b^r)/p}.
\end{equation}
In terms of the Fock PF operators, the number operator $n_p\equiv n_1$ counts the number of Fock PF associated with the couple $\alpha_1,\alpha_2$. The others PF are localized further to the left ($\alpha_0$) or to the right ($\alpha_3$) and their combined occupation is described by the number operator $n_0$.  Operatively, we can set $\hat{n}_p=(\hat{N}_L-\hat{N}_R-1)$ mod$(p)$ and  $\hat{n}_0+\hat{n}_p=(\hat{N}_b^l-\hat{N}_b^r-1)$ mod$(p)$. The parity operator for a set of sites $S$ is given by 
\begin{equation}
	P = (e^{i 2\pi/p})^{\sum_{i\in S} n_i}.
\end{equation}

The Fock space associated with two couples of $\mathbb{Z}_p$ PF is $p^2$-dimensional. If the total PF parity $P= (e^{i 2\pi/p})^{n_0+n_1}$ is conserved, the whole Fock space splits into $p$ decoupled sectors with a fixed total PF parity. In the following, we will work within one of these sectors, say the one with $P=1$. A convenient basis for this sector consists of the states $B=\{|n_p\rangle\}$ that are eigenstates of the number operator $\hat{n}_p |n_p\rangle = n_p |n_p\rangle$ with $n_p=0,\dots p-1$ (in what follows we will interchangeably use $n_1$ or $n_p$).

For the sake of concreteness, we now explicitly write the matrix representation of $d_1$ (and related operators) using the basis B. We focus, in particular, on the $p=6$ case. In particular, we have
\begin{eqnarray}\label{eq:ex6}
	d_1 &=& \left(\begin{array}{cccccc}
		0&1&0&0&0&0\\
		0&0&1&0&0&0\\
		0&0&0&1&0&0\\
		0&0&0&0&1&0\\
		0&0&0&0&0&1\\
		0&0&0&0&0&0\\
	\end{array}\right),
	\quad 
	n_1 = \left(\begin{array}{cccccc}
		0&0&0&0&0&0\\
		0&1&0&0&0&0\\
		0&0&2&0&0&0\\
		0&0&0&3&0&0\\
		0&0&0&0&4&0\\
		0&0&0&0&0&5\\
	\end{array}\right),\nonumber \\
	\alpha_1 &=&  \left(\begin{array}{cccccc}
		0&1&0&0&0&0\\
		0&0&1&0&0&0\\
		0&0&0&1&0&0\\
		0&0&0&0&1&0\\
		0&0&0&0&0&1\\
		1&0&0&0&0&0\\
	\end{array}\right).
\end{eqnarray}
From these expressions, it is evident that $\alpha_1$ cycles between the $6$ different occupation numbers defined by the two PF on the opposites sides of the weak-link. 

\section{Parafermionic Josephson effect}
The goal is to find the Hamiltonian describing the coupling between the two PF $\alpha_1$ and $\alpha_2$, localized on the two opposite sides of the weak link. Since this coupling corresponds to the tunneling of quasiparticles with fractional charge $e/p$ between the two superconductors, we expect the energy to be $2p\pi$-periodic in the phase difference $\varphi$ between the SCs. In Ref.\ \cite{Clarke2013} it has been derived the low-energy Hamiltonian
\begin{equation}
	H_{\rm PF} = E_{\rm PF} \left(\text{mod}\left[\frac{\varphi}{p}+\pi+\frac{2\pi n_1}{p}, 2\pi\right]-\pi\right)^2
\end{equation}
for a gapless weak link (i.e. no scattering between the two SC that gaps out the edge states) and ``even" $p=2m$ PF. Note the expected dependence on $\varphi/p$ which leads to the $2p\pi$ periodicity, as well as the phase shift depending on the number operator $n_1$. In presence of a strong gap in the weak-link, we describe the coupling between the PF as 
\begin{equation}
	H_{PF} = \gamma^{(1)} \cos\left(\frac{\varphi}{p} + \frac{2\pi n_1}{p} \right),
\end{equation}
that generalizes the Majorana-mediated ($p=2$)  $4\pi$ anomalous Josephson effect and features all the expected properties. Note that
\begin{equation}
	e^{-i \pi/p} \alpha_1^\dagger \alpha_2 = e^{i 2 \pi N_1/p},
\end{equation}
which can be easily verified by taking matrix elements $\langle n | e^{-i \pi/p} \alpha_1^\dagger \alpha_2 | m \rangle$ between Fock parafermion states $|n\rangle$.

\begin{figure}[t]
	\includegraphics[width=\linewidth]{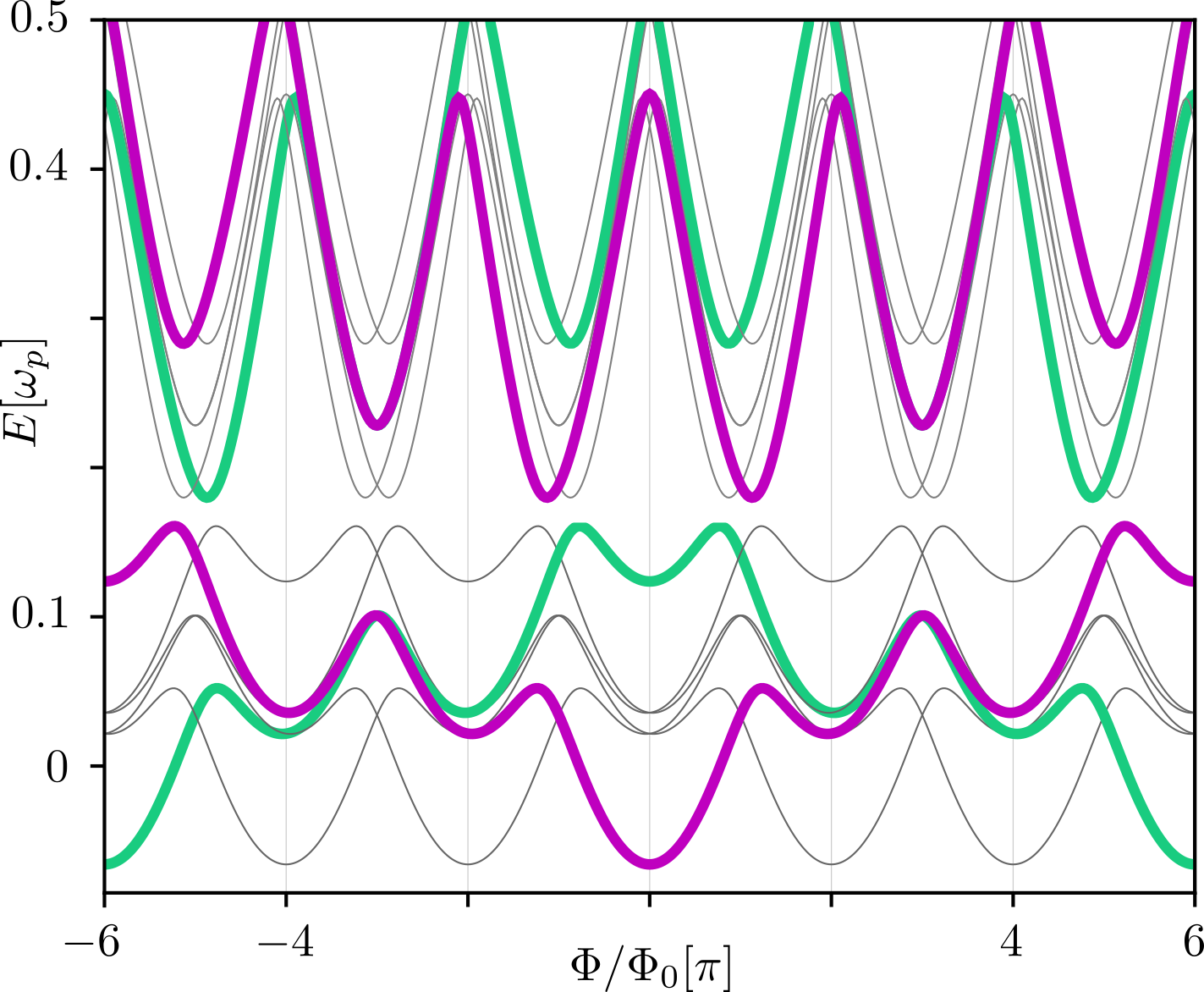}
	\caption{
		Spectrum of the whole system as a function of the external flux $\Phi$ for the case of $\gamma^{(1)} =0.07\,\omega_p$, $\gamma^{(2)}=0$ and $\gamma^{(3)}=0.035\,\omega_p$ and with fluxonium parameters given by $E_J=0.4\,\omega_p$, $E_L=0.03\,\omega_p$, $E_C=0.3125\,\omega_p$, with $\omega_p=\sqrt{8E_CE_J}$. Different lines refer to different values of $n_p$. In particular, the thick purple (green) lines highlights the $n_p=0$ ($n_p=3$)
		\label{Fig1-SM}
	}
\end{figure}

Higher order coupling at the junction is accounted for by noticing that $\left(e^{-i\pi/p} \alpha_1^\dagger \alpha_2 e^{i\varphi /p}\right)^k = e^{ik\varphi/p}e^{i2k \pi n_1/p}$ so that additional tunneling terms read
\begin{align}
	H_{\rm PF}^{(k)} &= \gamma^{(k)} \cos\left(
	\frac{k\varphi}{p}+\frac{2k\pi n_1}{p}\right).
\end{align}
If we focus on a given sector with fixed $n_1$, say $n_1=0$, the potential terms associated with the tunneling on the JJ reads
\begin{eqnarray}
	\label{eq:pot_all}
	U(\varphi)&=&E_J \cos(\varphi) + \gamma^{(1)}  \cos(\varphi/6)\nonumber\\
	& + & \gamma^{(2)} \cos(\varphi/3)  +  \gamma^{(3)} \cos(\varphi/2),
\end{eqnarray}  
and accounts for tunneling of: Cooper pairs, single PFs, pairs of PFs, $e$-charged quasiparticles (three PFs). As long as $\gamma\neq 0$, this part of the potential is thus $12\pi$-periodic in $\varphi$. Moreover, if there are no relative extra phases (which would come, e.g., from complex amplitudes) we have the additional symmetry $\varphi \leftrightarrow -\varphi$.  The spectrum with the additional term $\gamma^{(3)}$ on, describing electron tunneling at the junction, is shown in Fig.~\ref{Fig1-SM}, with the cases $n_p=0$ and $n_p=3$ highlighted as thick purple and green lines, respectively.

\begin{figure}[t]
	\includegraphics[width=\linewidth]{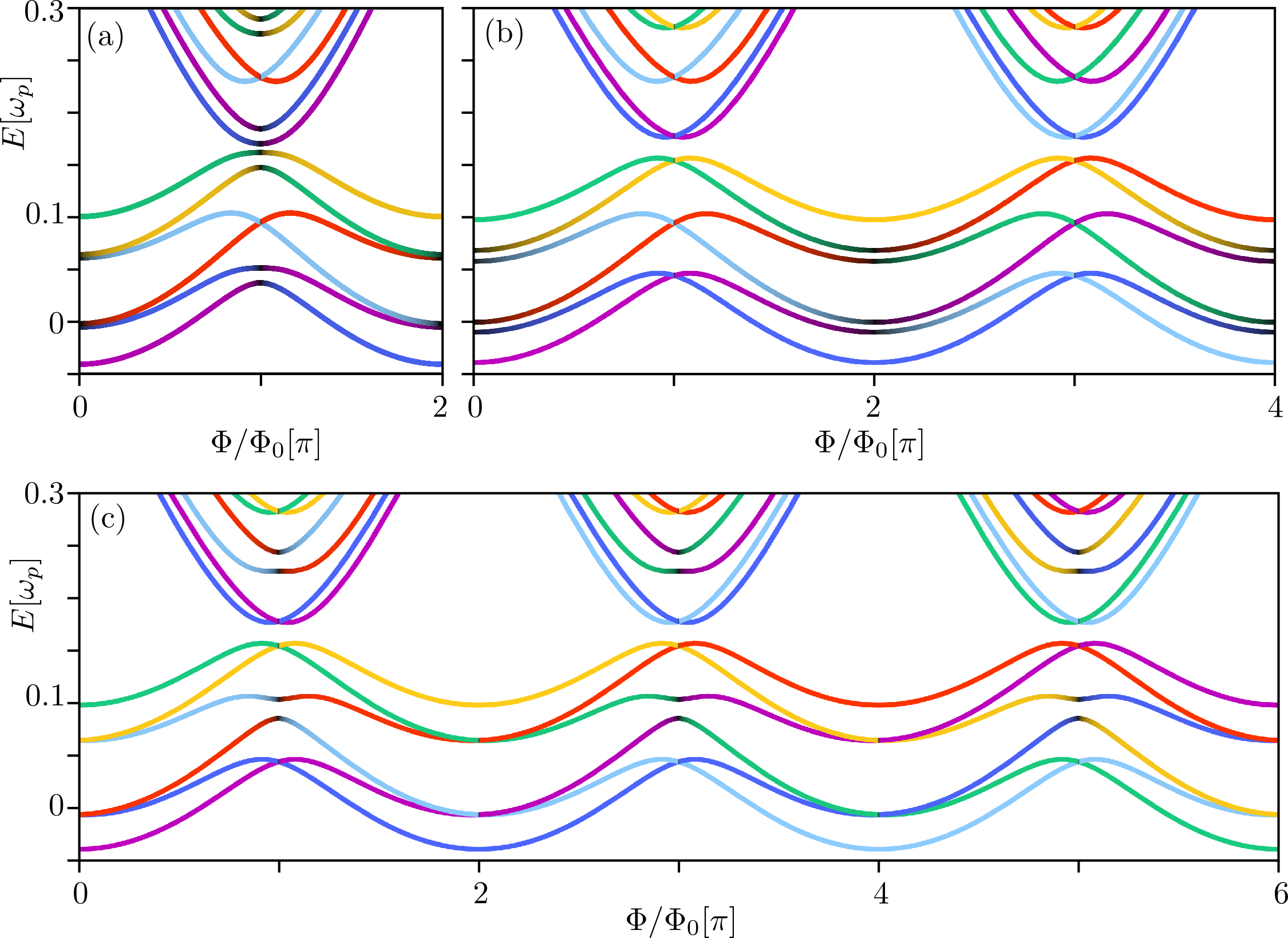}
	\caption{
		Periodicty of the spectrum of the whole system, for different off-diagonal terms $\lambda^{(k)}$. In (a), a finite $\lambda^{(1)}= 0.015\omega_p$ makes the spectrum $2\pi$-periodic. In (b), a finite $\lambda^{(2)}= 0.01\omega_p$ makes the spectrum $4\pi$-periodic. In (c), a finite $\lambda^{(3)}= 0.01\omega_p$ makes the spectrum $6\pi$-periodic. The other parameters are: $\gamma^{(1)} =0.07\,\omega_p$, $\gamma^{(2)}=\gamma^{(3)}=0$, $E_J=0.4\,\omega_p$, $E_L=0.03\,\omega_p$, $E_C=0.3125\,\omega_p$, with $\omega_p=\sqrt{8E_CE_J}$. The color scheme is the same as in Fig.2 of the main text. 
		\label{Fig2-SM}
	}
\end{figure}

\section{Parafermion couplings and spurious terms}

We now consider finite PF couplings, in terms of tunneling of quasiparticles, through the SC leads. By defining $\hat{V}=\alpha_0^\dag\alpha_1$ we have that $(\hat{V}^\dag)^k\hat{N}_L\hat{V}^k=\hat{N}_L+k$, so that tunneling of a charge $ke/m$ quasiparticles through the left SC lead is associated with $2k\pi $ phase slip of the $\phi$ field, whose amplitude $\lambda^{(k)}$ can be calculated through instanton technique analogously to the tunneling through the Hall droplet $\gamma^{(k)}$. In this case, the coherence length of the superconductor sets the scale of the process and in general in the perturbative regime the tunneling of $k$ quasiparticle of charge $e/m$ scales as $\lambda^{(k)}\propto\lambda^k$. 

Focusing on the left SC lead, the most general process is described by the generic term 
\begin{equation}
	H^{(1)}_{\rm PF}=\sum_{k,n=1}^3\lambda^{(k)}(\alpha^\dag_0)^k\alpha_1^n + {\rm H.c.}.
\end{equation}
By fixing the total parity we require $H^{(1)}_{\rm PF}$ to commute with $P$ and thus fix $n=k$. By further restricting our analysis on a sector with defined global PF parity, and thus on a $6$-dimensional space spanned by the basis $B$ consisting of the eigenstates of $n_p$, the generic coupling of PS $\alpha_0,\alpha_1$ can be effectively expressed as
\begin{equation}\label{Eq:H1PF}
	H^{(1)}_{\rm PF} = \sum_{k=1}^{p/2} \lambda^{(k)} \alpha_1^k + {\rm H.c.},
\end{equation}
where the $\lambda^{(k)}$'s can be complex. Note that, basically, the three matrices $\alpha_1$, $\alpha_1^2$ and $\alpha_1^3$ are the three generalization of the $\sigma_x$ matrix used for the Majorana case. We could also envision more complicated couplings that involve three or four PFs, while still conserving the total PF parity. Those terms would results in matrices entering Eq.~(\ref{Eq:H1PF}) that do not only contains ”1”s, but also complex phases. In this sense, they would represent a generalization of the $\sigma_y$ operator that could be considered in the Majorana case as a result, for example, of the coupling $\gamma_0\gamma_1\gamma_2\gamma_1=-\gamma_0\gamma_2$ (with $\gamma_i$ Majorana fermion operators, not to be confused with the amplitudes $\gamma^{(k)}$). Nevertheless, in general we expect the ``selection” rules induced by the spurious terms and described in the main text to still be valid. 

\subsection{Reduced periodicity}
As discussed in the main text, the coupling between the two PFs on the left SC lead determines a reduction of the periodicity of the spectrum as a function of the external magnetic flux. In particular, tunneling of charge $ke/3$ quasiparticles, with amplitude $\lambda^{(k)}$, makes the spectrum $2k\pi$-periodic. In Fig.~\ref{Fig2-SM}, we explicitly show this effect. In panel (a), obtained for a finite $\lambda^{(1)} = 0.015\,\omega_p$, all the level crossings are split and the spectrum is thus $2\pi$-periodic. Indeed, if we focus on the ground state, its adiabatic evolution when the flux is ramped from $\Phi/\Phi_0=0\to2\pi$ is still the ground state of the system. The same applies for all the other excited states, since all the crossings are lifted. The situation is different in panel (b), where we consider only a finite $\lambda^{(2)} = 0.01\,\omega_p$. There, focusing again on the ground state, the presence of protected crossings requires the flux to be (adiabatically) ramped all the way to $\Phi/\Phi_0=0\to4\pi$ in order to recover the initial situation. The same applies to all the other levels and the spectrum is thus $4\pi$ periodic. Finally, in panel (c) we consider the effect of a finite $\lambda^{(3)} = 0.01\,\omega_p$. In this case, even more crossings are protected and the flux has to be increased from $\Phi/\Phi_0=0\to6\pi$ to recover the initial configuration. The periodicity is thus $6\pi$. 

\begin{figure}[t]
	\includegraphics[width=\linewidth]{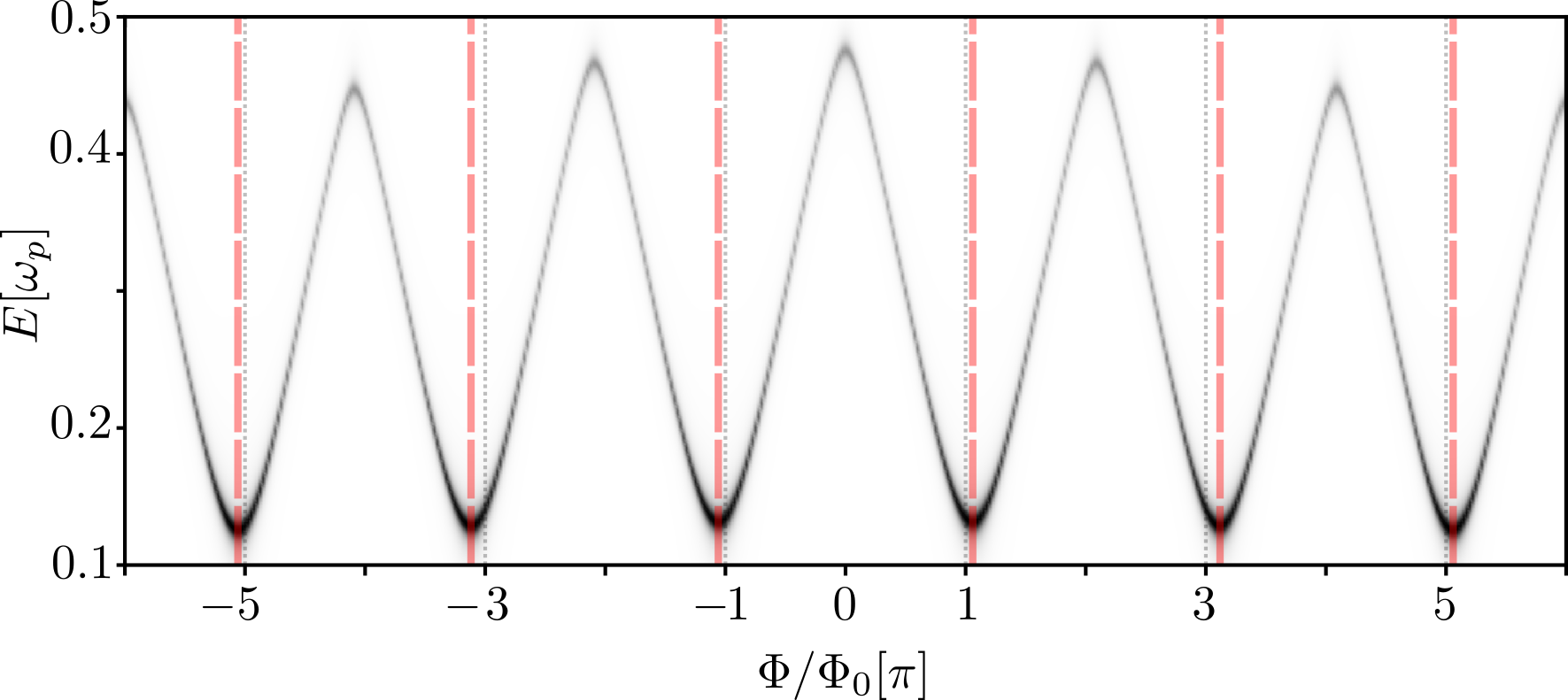}
	\caption{
		MWS of the system as a function of the external flux $\Phi$ for the entire $12\pi$ periodicity, for the $n_p=0$ case and for $\gamma=0.07~ \omega_p$, $\gamma^{(2)}=\gamma^{(3)}=0$ and with fluxonium parameters given by $E_J=0.4~\omega_p$, $E_L=0.03~\omega_p$, $E_C=0.3125~\omega_p$, with $\omega_p=\sqrt{8E_CE_J}$. The vertical red dashed lines indicate the positions of the minima, which differ from integer multiple of $\pi$ (shown as vertical dotted lines), according to the analytical approximation for $\Phi_m^{(n_p=0)}$, given in Eq.~\eqref{eq:PhiM}.
		\label{Fig3-SM}
	}
\end{figure}

\section{Microwave spectrum}

\begin{figure}[t]
	\includegraphics[width=.9\linewidth]{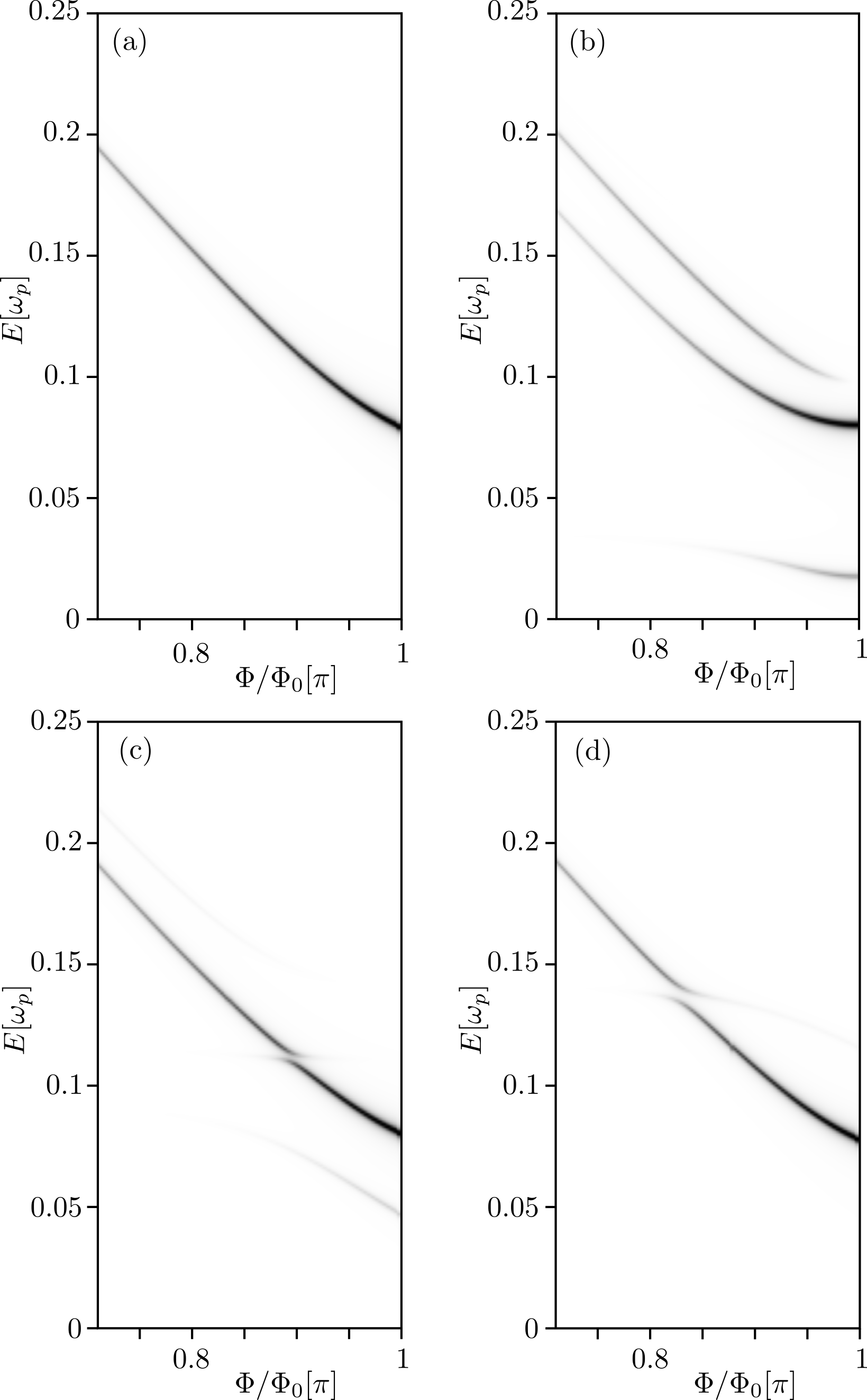}
	\caption{
		MWS of the system as a function of the external flux $\Phi$ without off-diagonal couplings [panel (a)], with $\lambda^{(1)}=0.02\omega_p$ [panel (b)], with $\lambda^{(2)}=0.02~\omega_p$ [panel (c)], and with $\lambda^{(3)}=0.02~\omega_p$ [panel (d)]. All the other parameters are the same as in Fig. 3 of the main text, i.e. $E_L=0.03\,\omega_p,\,E_J=0.5\,\omega_p,\,\gamma=0.05\,\omega_p$, and $\lambda^{(1)}=\lambda^{(2)}=\lambda^{(3)}=0.02~\omega_p$.
		\label{Fig4-SM}
	}
\end{figure}

Assuming the system to be in the state $|n\rangle$, with energy $\omega_n$ (which can be the ground state or another state of the lowest band, for example with a fixed number of PF), the amplitude $I_{n\to m}$ associated with the microwave-induced transition to the state $|m\rangle$ with energy $E_m>E_n$ is computed through the Fermi golden rule expression 
\begin{equation}
	S_{n\to m}(\omega) = \left|\langle \psi_n |\hat \phi| \psi_m \rangle\right|^2\delta(\omega-\omega_{nm}).
\end{equation}
with $\omega_{nm}=\omega_m-\omega_n$.

\subsection{Position of the minima of the microwave spectrum (no spurious terms)}
Without additional PF couplings, neither diagonal ($\gamma^{(2)} = \gamma^{(3)} = 0 $) nor off-diagonal ($\lambda^{(k)} = 0$), it is possible to analytically determine the (approximate) positions of the minima of the microwave spectrum (MWS). The latter are associated with anticrossings in the fluxonium spectrum and, therefore, occur whenever the $\varphi$-dependent potential energy $V(\varphi)$ of the whole system features two equal global minima. We have
\begin{equation}
	\frac{V(\varphi)}{E_J} = \frac{E_L}{2E_J} (\varphi-\Phi/\Phi_0)^2 - \cos(\varphi) - \frac{\gamma}{E_J} \cos\left(
	\frac{\varphi}{6} + \frac{2\pi n_1}{6}\right).
\end{equation}
For $\gamma=0$, it is easy to verify that the condition for having two equal global minima is $\Phi_k/\Phi_0=(2k-1) \pi$ ($k\in \mathbb{Z}$). Let us study how this scenario is modified in presence of a finite $\gamma$, but assuming the regime $E_J \gg  \gamma,E_L$. In this case, the position of the minima is still mainly controlled by the cosine with prefactor $1$. By expanding the potential around the different minima of this cosine, one can find the condition 
\begin{eqnarray}
	\label{eq:PhiM}
	&&\frac{\Phi_m^{(n_1)}}{\Phi_0} = (2m+1)\pi \\
	&&\quad+ \frac{\gamma}{E_L} \frac{	\cos[(n_1+m) \pi /3]-\cos[(n_1+m+1)\pi /3]}{2\pi^2}, \nonumber
\end{eqnarray}
specifying the (approximate) values of the flux associated with minima in the MWS.

In the MW spectrum, the $12\pi$ periodicity mainly manifest itself in terms of the position of the minima, which deviates from $(2k-1)\pi$, where the amplitude of these deviations is proportional to $\gamma/E_L$ for $\gamma,E_L\ll E_J$. The full MW spectrum showing the $12\pi$ periodicity is shown in Fig.~\ref{Fig3-SM}: the position of the minima is indeed slightly shifted from multiples of $\Phi/\Phi_0=\pi$. In turn, the maxima also show a marked frequency dependence. Position and energy of the minima and maxima clearly depend on the overall energy-phase relation

\subsection{Effects of off-diagonal couplings}
In Fig.~3 of the main text, we chose the parameters so that some states belonging to the second fluxonium band intersect with the ones of the first band. This leads to additional crossings that make it straightforward to detect and distinguish the presence of different non-zero $\lambda^{(k)}$ from the analysis of the MWS. In particular, to show the maximal effect of PF hybridization, there we turned on all the $\lambda^{(k)}$. Here, in Fig.~\ref{Fig4-SM}, we consider the same scenario, but we consider only a single non-vanishing $\lambda^{(k)}$ at a time. In panel (a), all $\lambda^{(k)}$ are set to zero and no avoided crossing appears in the MWS spectrum. For a finite $\lambda^{(1)}=0.02~\omega_p$, in panel (b), a large splitting at $\Phi/\Phi_0=\pi$ is observed. For a finite $\lambda^{(2)}=0.02~\omega_p$, in panel (c), a clear splitting appears around $\Phi/\Phi_0\sim 0.9~\pi$. Finally, for a finite $\lambda^{(3)}=0.02~\omega_p$, in panel (d) a clear splitting appears around $\Phi/\Phi_0\sim 0.82~\pi$. This analysis allows us to claim that each of the three big splittings shown in Fig. 3 (b) of the main text, and highlighted with three colored arrows, is associated with a specific coupling term $\lambda^{(k)}$.


\begin{thebibliography}{100}
\bibitem{noteEJ}
We notice that ordinary tunneling of Cooper pairs between the SC leads can be highly suppressed by the quantum Hall gap. Nevertheless, we can always add a second Josephson junction in parallel with the superinductor outside of the Hall bar, so to have a well defined fluxonium that does not depend on the details of the parafermionic junction.
\end{thebibliography}
\end{document}